\documentclass[11pt]{article}
\usepackage{amsmath,color}
\usepackage{amssymb}
\usepackage{braket}
\usepackage{acro}
\usepackage{orcidlink}

\def\dd{\mathrm d}
\def\lag{\mathcal L}
\def\tr{\mathrm{tr}}
\DeclareAcronym{cft}{
    short = CFT,
    long = conformal field theory
}
\DeclareAcronym{fg}{
    short = FG,
    long = Fefferman-Graham
}

\begin{document}

\begin{center}
    {\Large\bf Holographic Weyl Anomaly in 8d\\from General Higher Curvature Gravity}

    \bigskip

    Fei-Yu Chen\orcidlink{0000-0002-8674-9316}\textsuperscript{1} and H.~L\"u\orcidlink{0000-0001-7100-2466}\textsuperscript{1,2}

    \bigskip

    \textsuperscript{1}{\it Center for Joint Quantum Studies and Department of Physics,\\
    School of Science, Tianjin University, Tianjin 300350, China}

    \bigskip

    \textsuperscript{2}{\it Joint School of National University of Singapore and Tianjin University,\\
    International Campus of Tianjin University, Binhai New City, Fuzhou 350207, China}
\end{center}

\begin{abstract}
    We calculate the holographic central charges for general higher curvature gravity theory dual to eight dimensional CFT. To do this, we first elaborate the general form of Weyl anomaly in 8d CFT and find 11 non-trivial linearly independent curvature combinations, one of which is Euler density and the rest are Weyl invariants, including 7 non-differentiated ones and 3 differentiated ones. The Weyl invariants are constructed as invariant polynomials of curvature tensor and covariant derivatives. We denote $W_{(n)}$ as the Weyl invariant that contains a polynormial term with a minimum of $n$ curvature tensors. Interestingly, since there are a total of 12 Weyl invariants in 8d, our finding means two of them are trivial and expressible as total derivatives. The resulting central charges are expressed in terms of 15 theory-dependent constants. Remarkably, we find that the $W_{(2)}$ invariant corresponds to the $c$-charge that is proportional to $C_T$, while the two $W_{(3)}$'s are related to three-point function parameters of energy-momentum tensor. This suggests a possible connection between the $c$-charges of $W_{(n)}$'s and the $n$-point functions of energy-momentum tensor.
\end{abstract}

\newpage

\section{Introduction}
\label{sec:intro}
Energy-momentum tensor is an important object to study in quantum field theory as it's the most universal operator. In \ac{cft}, although the action is conformally invariant at classical level, it may be violated at quantum level in even-dimensional curved background and gives so-called Weyl anomaly, in which time the trace of energy-momentum tensor has a non-zero vacuum expectation value, whose form can be generally written as \cite{Duff:1977ay,Duff:1993wm}
\begin{equation}\label{eq:trace-anom-form}
    (4\pi)^{d/2}\braket{T_a^a} = -a E^{(d)} + \sum_i c_i I^{(d)}_i + \nabla_a J^a,
\end{equation}
where $E^{(d)}$ is the $d$-dimensional Euler density and $I^{(d)}_i$ are Weyl invariants, they are called type A and type B Weyl anomaly respectively. The third term $\nabla_a J^a$ is trivial anomaly, it's irrelevant since it can be removed by adding a local counter term. The coefficients $a$ and $c_i$ are known as $a$-charge and $c$-charges respectively, they are important characteristics of a given \ac{cft}. The number of $c$-charges increases with space-time dimension, we have 0, 1, 3 $c$-charges in 2d, 4d, and 6d. The $a$-charge is the $a$-function evaluated at fixed point \cite{Zamolodchikov:1986gt,Komargodski:2011vj} and, more interestingly, coincides with the holographic entanglement entropy in even dimensions\cite{Dowker:2010yj,Casini:2011kv}. In 4d the $c$-charge is proportional to the flat background energy-momentum tensor two-point function parameter $C_T$ \cite{Osborn:1993cr,Erdmenger:1996yc}
\begin{equation}
    \braket{T^{ab}(x)T^{cd}(y)} = C_T \frac{\mathcal I^{abcd}(x - y)}{|x - y|^{2d}},
\end{equation}
where $\mathcal I^{abcd}(x - y)$ is a tensor structure whose form is irrelavent here. In higher dimensions $C_T$ is given by some linear combination of the $c$-charges, one may recombine the Weyl invariants so that $C_T$ becomes proportional to one of the $c$-charges. It's then natural to ask which combination of the Weyl invariants corresponds to $C_T$. In 6d we have 3 Weyl invariants with $C_T$ corresponding to the one with one covariant derivative, the other two contains no covariant derivative. So we may conjecture that $C_T$ corresponds to the Weyl invariant with highest derivative order. To see this we have to consider higher dimensions such as 8d and 10d, in this work we'll consider the 8d case.

We do this in the framework of AdS/CFT correspondence, where the large $N$ limit of strongly-coupled \ac{cft} can be studied through weakly-coupled gravity theory in anti de-sitter (AdS) background \cite{Maldacena:1997re,Gubser:1998bc,Witten:1998qj}. As an extension to Einstein gravity, higher curvature gravities is insightful to study as they capture various new features of the dual \ac{cft}. They can be regarded as a tool to consider finite $N$ corrections \cite{Buchel:2004di,Buchel:2008sh,Buchel:2008ae}, or introducing new couplings between the operators in \ac{cft} \cite{Hofman:2008ar,Hofman:2009ug}.

The most important feature of higher curvature gravity is that the equation of motion contains at most four derivatives with respect to the metric, so the same is true for linearized equation of motion around AdS background. This introduces two extra modes to the graviton spectrum, the massive scalar mode and ghost-like spin-2 mode \cite{Stelle:1976gc,Bueno:2016ypa}. The spin-2 mode breaks unitarity of the dual \ac{cft} thus it's mandatory to decouple it, while the decoupling of the scalar mode is required by the holographic $a$-theorem. Theories where both modes are \textit{exactly} decoupled is known as massless gravity \cite{Li:2017ncu}. For perturbative higher curvature gravity such extra modes are decoupled perturbatively. The holographic aspects of higher curvature gravity has been studied extensively, \textit{e.g.} refs.\cite{Buchel:2009sk,Myers:2010jv,Bueno:2018xqc,Li:2019auk,Camanho:2013pda}.

The Weyl anomaly and the associated central charges, can be calculated holographically using \ac{fg} expansion \cite{Henningson:1998ey}. For higher curvature gravity, this method has been applied to the study of Gauss-Bonnet gravity \cite{Buchel:2009sk} and cubic quasi-topological gravity \cite{Myers:2010jv}. For more general higher curvature gravities, however, the calculation becomes extremely challenging. A much simpler approach known as reduced \ac{fg} expansion was later proposed \cite{Bugini:2016nvn,Li:2017txk}, and applied to various 4d and 6d cases \cite{Li:2021jfh,Li:2019auk,Lu:2019urr}. However, such method fails to apply for $d \geqslant 8$ since the number of $c$-charges is too large and it's not possible to find enough equations to solve all of them. So for 8d we have to apply the conventional \ac{fg} expansion method. Another challenge of this work is elaborating the general form of 8d Weyl anomaly as it's not present in the literature. At bulk side, instead of consider specific Lagrangian of gravity theory, we find that the \ac{fg} expansion can be done once-for-all for general higher curvature gravities instead of considering specific Lagrangians. The results are expressed in terms of a set of theory-dependent constants. We find some of the $c$-charges are expressible in terms of $C_T, C_T t_2, C_Tt_4$, where $t_2$ and $t_4$ are the energy flux parameters, they are directly related to the three-point function of energy-momentum tensor. The results also allow us to further classify the Weyl invariants in 8d by looking at their powers in curvature. We believe this relation can be generalized to arbitrary even dimensions so that the $c$-charges of $n$-th power Weyl invariants are related to the $n$-point function of the energy-momentum tensor.

The paper is organized as follows. In section \ref{sec:method}, we explain in detail the method we use to carry out the calculation. Since the calculation relies on the explicit forms of 8d Weyl anomalies, we derive and classify them using Weyl invariants in section \ref{sec:8d-weyl-anoms}. We present our results in section \ref{sec:result}, and conclude our work in section \ref{sec:conclusion}.

\section{Methodology}
\label{sec:method}

\subsection{Higher curvature gravities}
We begin by a brief review of higher curvature gravity, whose Lagrangian is a function of the metric and non-differentiated Riemann tensor. The general form of higher curvature gravity action can be written as
\begin{equation}
    S = \int\dd^D x \sqrt{|g|}\lag(g_{\mu\nu}, R_{\mu\nu\rho\sigma}).
\end{equation}
The equation of motion can be obtained by varying the action
\begin{equation}\label{eq:ghc-eom}
    \mathcal E^{\mu\nu} = -\frac{1}{\sqrt{|g|}}\frac{\delta S}{\delta g_{\mu\nu}} = P^{\mu\rho\sigma\gamma} R^{\nu}{}_{\rho\sigma\gamma} - \frac12\lag g^{\mu\nu} - 2\nabla_\rho\nabla_\sigma P^{\mu\rho\sigma\nu}, \qquad P^{\mu\nu\rho\sigma} = \frac{\partial\lag}{\partial R_{\mu\nu\rho\sigma}}.
\end{equation}
The tensor $P^{\mu\nu\rho\sigma}$ should inherits all algebraic properties of the Riemann tensor, \textit{i.e.}
\begin{equation}\label{eq:p-tensor-props}
    P^{\mu\nu\rho\sigma} = -P^{\nu\mu\rho\sigma} = -P^{\mu\nu\sigma\rho} = P^{\rho\sigma\mu\nu}, \qquad P^{\mu[\nu\rho\sigma]} = 0.
\end{equation}
In order to classify how the Riemann tensor enters the Lagrangian, it is useful to treat the Riemann tensor as an independent quantity of the metric. We introduce ``curvature-like'' quantities of the form $\hat R_{\mu\nu\rho\sigma} = 2\lambda g_{\mu[\rho}g_{\sigma]\nu}$, where the metric $g_{\mu\nu}$ may not necessarily be the metric of maximally symmetric space. In other words, $\hat R_{\mu\nu\rho\sigma}$ satisfies the above algebraic relations of the Riemman tensor, but not the usual different relation of the metric.

When evaluating on such hatted curvature, $P^{\mu\nu\rho\sigma}$ can only be built from the metric $g_{\mu\nu}$, its form is uniquely determined by \eqref{eq:p-tensor-props} up to a constant factor \cite{Bueno:2016xff}
\begin{equation}\label{eq:lagpdr1-form}
    \hat P^{\mu\nu\rho\sigma} \equiv P^{\mu\nu\rho\sigma}(g_{\mu\nu}, \hat R_{\mu\nu\rho\sigma}) = k_{1,1}g^{\mu[\rho}g^{\sigma]\nu}.
\end{equation}
We can further consider general higher derivatives of the Lagrangian, \textit{i.e.}
\begin{equation}
    P_n^{\mu_1\nu_1\rho_1\sigma_1\cdots\mu_n\nu_n\rho_n\sigma_n} = \frac{\partial^n\lag}{\partial R_{\mu_1\nu_1\rho_1\sigma_1}\cdots\partial R_{\mu_n\nu_n\rho_n\sigma_n}},
\end{equation}
which have similar properties as \eqref{eq:p-tensor-props} for each group of indices corresponding to a Riemann tensor, plus the invariance under any permutation between the index groups. The form of such tensors when evaluated on $\hat R_{\mu\nu\rho\sigma}$ are also fixed, and enumerating their tensor structures is equivalent to enumerating Riemann scalars of order $n$. To see this, we note that the tensor $P_n$ is \textit{defined} by $\delta^n\lag$ through
\begin{equation}
    \delta^n\lag = P_n^{\mu_1\nu_1\rho_1\sigma_1\cdots\mu_n\nu_n\rho_n\sigma_n} \delta R_{\mu_1\nu_1\rho_1\sigma_1}\cdots\delta R_{\mu_n\nu_n\rho_n\sigma_n}.
\end{equation}
When evaluated on $\hat R_{\mu\nu\rho\sigma}$, $\delta^n\lag$ is given by the linear combination of all possible scalars built from $\delta R_{\mu\nu\rho\sigma}$, which has the same symmetry properties as the Riemann tensor. Thus the general form of $P_n$ when evaluated on $\hat R_{\mu\nu\rho\sigma}$ can be chosen as
\begin{equation}\label{eq:lagpdr-general-def}
    \hat P_n^{\mu_1\nu_1\rho_1\sigma_1\cdots\mu_n\nu_n\rho_n\sigma_n} = \sum_i k_{n,i}\,\frac{\partial^n\mathcal R^{(n)}_i}{\partial R_{\mu_1\nu_1\rho_1\sigma_1}\cdots\partial R_{\mu_n\nu_n\rho_n\sigma_n}},
\end{equation}
where $\mathcal R^{(n)}_i$ denotes the $i$-th scalar built from the contraction of $n$ Riemann tensors, $k_{n,i}$ are constants. In this work we need the lists of $\mathcal R^{(n)}_i$ for $n \leqslant 4$. For $n=1,2,3$, we have
\begin{align}
    \mathcal R^{(1)} & = \{R\},\nonumber\\
    \mathcal R^{(2)} & = \{R^2, R_{\mu\nu}R^{\mu\nu}, R_{\mu\nu\rho\sigma}R^{\mu\nu\rho\sigma}, \Box R\},\nonumber\\
    \mathcal R^{(3)} & = \{R^3,RR^{\mu\nu}R_{\mu\nu},R^{\mu\nu}R_{\mu}{}^{\rho}R_{\nu\rho},R_{\mu\rho}R_{\nu\sigma}R^{\mu\nu\rho\sigma},RR^{\mu\nu\rho\sigma}R_{\mu\nu\rho\sigma},R_{\sigma\alpha}R^{\mu\nu\rho\sigma}R_{\mu\nu\rho}{}^{\alpha},\nonumber\\
    & \qquad R^{\mu\nu\rho\sigma}R_{\mu\nu}{}^{\alpha\beta}R_{\rho\sigma\alpha\beta},R^{\mu\nu\rho\sigma}R_{\mu}{}^{\alpha}{}_{\rho}{}^{\beta}R_{\nu\alpha\sigma\beta},\nabla^{\mu}R\nabla_{\mu}R,\nabla^{\rho}R^{\mu\nu}\nabla_{\rho}R_{\mu\nu},\nabla^{\rho}R^{\mu\nu}\nabla_{\nu}R_{\mu\rho},\nonumber\\
    & \qquad \nabla^{\alpha}R^{\mu\nu\rho\sigma}\nabla_{\alpha}R_{\mu\nu\rho\sigma},R\Box R,\nabla^{\nu}\nabla^{\mu}RR_{\mu\nu},\Box R^{\mu\nu}R_{\mu\nu},\nabla_{\sigma}\nabla_{\nu}R_{\mu\rho}R^{\mu\nu\rho\sigma},\Box^2R\}.\label{eq:r123-list}
\end{align}
Note that in addition to the pure polynomials of the Riemman tensor, we also listed above the terms involving covariant derivatives, which also serves the basis for calculating Weyl invariants and Weyl anomalies in the later section. The basis for $\mathcal R^{(4)}$ is more lengthy, we have totally 26 non-differentiated terms, plus 66 derivative terms, which we give in \eqref{eq:r4-list} of appendix \ref{sec:ap:reimann-scalars}. Since the order of the derivative is also $n$, the right hand side of the above expression only consist of the metric $g^{\mu\nu}$ and no Riemann tensors. Note that when taking the derivative with respect to the Riemann tensor, one should impose the properties \eqref{eq:p-tensor-props}, \textit{i.e.}
\begin{equation}
    \frac{\partial R_{\alpha\beta\gamma\delta}}{\partial R_{\mu\nu\rho\sigma}} = T_{\alpha\beta\gamma\delta}^{\mu\nu\rho\sigma} - T_{\alpha\beta\gamma\delta}^{\mu[\nu\rho\sigma]}, \qquad T_{\alpha\beta\gamma\delta}^{\mu\nu\rho\sigma} = \frac12\left(\delta_\alpha^{[\mu}\delta_\beta^{\nu]}\delta_\gamma^{[\rho}\delta_\delta^{\sigma]} + \delta_\gamma^{[\mu}\delta_\delta^{\nu]}\delta_\alpha^{[\rho}\delta_\beta^{\sigma]}\right).
\end{equation}
For some low-lying examples, we consider the general forms of $\hat P_1 \equiv \hat P$ and $\hat P_2$. The corresponding Riemann curvature invariants are
\begin{equation}
    \mathcal R^{(1)} = \{R\}, \qquad \mathcal R^{(2)} = \{R^2, R_{\mu\nu}R^{\mu\nu}, R_{\mu\nu\rho\sigma}R^{\mu\nu\rho\sigma}\},
\end{equation}
thus we have
\begin{align}
    \hat P^{\mu\nu\rho\sigma} & = k_{1,1}\frac{\partial R}{\partial R_{\mu\nu\rho\sigma}} = 
    k_{1,1}g^{\mu[\rho}g^{\sigma]\nu},\cr
    \hat P_2^{\mu\nu\rho\sigma\alpha\beta\gamma\delta} & = k_{2,1}\frac{\partial^2 R^2}{\partial R_{\mu\nu\rho\sigma}\partial R_{\alpha\beta\gamma\delta}} + k_{2,2}\frac{\partial^2 (R_{\mu\nu}R^{\mu\nu})}{\partial R_{\mu\nu\rho\sigma}\partial R_{\alpha\beta\gamma\delta}} + k_{2,3}\frac{\partial^2 (R_{\mu\nu\rho\sigma}R^{\mu\nu\rho\sigma})}{\partial R_{\mu\nu\rho\sigma}\partial R_{\alpha\beta\gamma\delta}}\nonumber\\
    & = C^{\mu\nu\rho\sigma\alpha\beta\gamma\delta} - C^{\mu[\nu\rho\sigma]\alpha\beta\gamma\delta} - C^{\mu\nu\rho\sigma\alpha[\beta\gamma\delta]} + C^{\mu[\nu\rho\sigma]\alpha[\beta\gamma\delta]},\label{eq:lag-pd-r-2-form}
\end{align}
where
\begin{align}
    C^{\mu\nu\rho\sigma\alpha\beta\gamma\delta} & = 2k_{2,1}g^{\mu[\rho}g^{\sigma]\nu}g^{\alpha[\gamma}g^{\delta]\beta} + 2k_{2,2}\delta_{(\epsilon}^{[\mu} g^{\nu]}\delta_{\kappa)}^{[\rho} g^{\sigma]}g^{\epsilon[\alpha}g^{\beta][\gamma}g^{\delta]\kappa}\nonumber\\
    & \qquad + k_{2,3}\left(g^{\mu[\alpha}g^{\beta]\nu}g^{\rho[\gamma}g^{\delta]\sigma} + g^{\mu[\gamma}g^{\delta]\nu}g^{\rho[\alpha}g^{\beta]\sigma}\right).
\end{align}
The last line in \eqref{eq:lag-pd-r-2-form} ensures the Bianchi identity of $\hat P$ inherited from the Riemann tensor so that the $C$-tensor given above is more compact. These results were obtained in ref.~\cite{Bueno:2016ypa} with a different parametrization, where parameters $(e, a, b, c)$ are used instead and related to our parameters by
\begin{equation}
    k_{1,1} = 2e, \qquad k_{2,1} = 2b, \qquad k_{2,2} = 2c, \qquad k_{2,3} = a.
\end{equation}

Now we return to general higher curvature gravity theory. We assume the theory admits maximally symmetric solution $\bar g_{\mu\nu}$ with curvature $\bar R_{\mu\nu\rho\sigma} = 2\lambda\bar g_{\mu[\rho}\bar g_{\sigma]\nu}$, which has the same form as $\hat R_{\mu\nu\rho\sigma}$ with $g_{\mu\nu}$ replaced by $\bar g_{\mu\nu}$. We can then substitute $\hat P$ into \eqref{eq:ghc-eom} to obtain
\begin{equation}\label{eq:lag-vac-sol}
    \bar{\mathcal E}^{\mu\nu} = \bar g^{\mu\nu}\left[(D - 1)\lambda k_{1,1} - \frac12\bar\lag\right] \quad \implies \quad \bar\lag = 2(D - 1)\lambda k_{1,1}.
\end{equation}
The corresponding linearized theory can be obtained by perturb \eqref{eq:ghc-eom} around maximally symmetric space. Doing this introduces the tensor $\bar P_2$ and the related parameters depend on $k_{2,i}$. The spectrum contains extra scalar mode and ghost-like spin-2 mode, with masses $m_s$ and $m_g$. The linearized equation of motion can be casted into the form
\begin{equation}
    \bar{\mathcal E}^{\mu\nu}_L = \frac{1}{\kappa_{\rm eff}}(\Box - 2\lambda)\delta g^{\mu\nu} + \cdots,
\end{equation}
where ellipsis denotes contribution from the two extra modes, $\kappa_{\rm eff}$ is the effective Einstein gravity constant. These parameters are given by (after adapting to our parametrization) \cite{Bueno:2016ypa}
\begin{align}
    \kappa_{\rm eff}^{-1} & = 2k_{1,1} - 8(D - 3)\lambda k_{2,3},\\
    m_g^2 & = \frac{-k_{1,1} + 4 (D-3) \lambda k_{2,3}}{k_{2,2} + 4 k_{2, 3}},\\
    m_s^2 & = \frac{4\lambda\left(D^2 k_{2,1}+D \left(k_{2,2}-k_{2,1}\right)-k_{2,2}+2 k_{2,3}\right)-(D-2) k_{1,1}}{4 (D-1) k_{2,1}+D k_{2,2}+4 k_{2,3}}.
\end{align}
The two-point function parameter $C_T$ is directly related to $\kappa_{\rm eff}$ \cite{Bueno:2018yzo}
\begin{equation}\label{eq:ct-from-kappaeff}
    C_T = \frac{\Gamma(d + 2)}{\pi^{d/2}(d - 1)\Gamma(d/2)}\frac{L^{d - 1}}{\kappa_{\mathrm{eff}}}\,.
\end{equation}
The massless conditions then requires $m_s$ and $m_g$ to infinity, or sending their denominators to zero. For the perturbative approach to higher curvature gravity, the parameters $k_{2,i}$ are infinitesimal, so the two extra modes are decoupled at leading order. In our calculation, we find it not necessary to impose the massless conditions explicitly, so our results are valid both for finite coupling approach after imposing the massless conditions, or perturbative approach to higher curvature gravity.

\subsection{Holographic Weyl anomaly}
Now we turn to the calculation of holographic Weyl anomaly. This is done by evaluating the action on asymptotically locally anti-de Sitter space, which is described by the \ac{fg} metric
\begin{equation}\label{eq:fg-expansion}
    \dd s^2 = \frac{L^2}{4\rho^2}\dd\rho^2 + \frac{1}{\rho}g_{ab}\dd x^a\dd x^b,
\end{equation}
where $g_{ab}(\rho)$ is the boundary metric. Near the boundary $\rho = 0$ the metric has the following asymptotic expansion
\begin{equation}\label{eq:fg-metric-expand}
    g_{ab}(\rho) = g_{ab}^{(0)} + \rho g_{ab}^{(1)} + \cdots + \rho^{d/2} g_{ab}^{(d/2)} + \rho^{d/2} \ln\rho \, h_{ab}^{(d/2)} + \mathcal O(\rho^{d/2 + 1}).
\end{equation}
The action can be expanded accordingly
\begin{equation}
    S = \int\dd^d x\int\dd\rho \frac{L}{2\rho^{1 + d/2}}\sqrt{|g|}\lag = \frac{L}{2}\int\dd^d x\sqrt{|g^{(0)}|}\int\dd\rho \frac{1}{\rho^{1 + d/2}}(a_0 + \rho a_1 + \cdots + a_{d/2}\rho^{d/2} + \cdots).
\end{equation}
The Weyl anomaly is given by \cite{Henningson:1998ey}
\begin{equation}
    \braket{T_a^a} = L a_{d/2}.
\end{equation}
This term yields a logarithm after the integration, which breaks conformal invariance.

Under the \ac{fg} expansion \eqref{eq:fg-expansion}, the Riemann tensor has the form
\begin{equation}\label{eq:fg-riemann-form}
    R_{\mu\nu\rho\sigma} = -\frac{2}{L^2}g_{\mu[\rho}g_{\sigma]\nu} + \Delta R_{\mu\nu\rho\sigma} \equiv \hat R_{\mu\nu\rho\sigma} + \Delta R_{\mu\nu\rho\sigma},
\end{equation}
where the non-zero components of $\Delta R_{\mu\nu\rho\sigma}$ are given by\footnote{$\hat\rho$ is the index of the $\rho$ coordinate, other free indices are all boundary indices. All fields on the right hand side are boundary fields, and indices are raising and lowering using the boundary metric. $\prime = \partial_\rho$.}
\begin{subequations}\label{eq:delta-r-comps}
    \begin{align}
        \Delta R_{\hat\rho a\hat\rho b} & = \frac{1}{4\rho}\left(g'_a{}^c g_{bc} - g''_{ab}\right),\\
        \Delta R_{\hat\rho abc} & = \frac{1}{\rho}\nabla_{[b}g'_{c]a}\,,\\
        \Delta R_{abcd} & = \frac{2}{L^2\rho}g_{a[c}g'_{d]b} - \frac{2}{L^2}g'_{a[c}g'_{d]b} + \frac{1}{\rho} R_{abcd}\,.
    \end{align}
\end{subequations}
We can then expand $\lag$ with respect to $\Delta R_{\mu\nu\rho\sigma}$
\begin{equation}\label{eq:lag-r-expand}
    \lag(g_{\mu\nu}, R_{\mu\nu\rho\sigma}) = \lag_0 + \hat P^{\mu\nu\rho\sigma}\Delta R_{\mu\nu\rho\sigma} + \frac{1}{2!}\hat P_2^{\mu\nu\rho\sigma\alpha\beta\gamma\delta}\Delta R_{\mu\nu\rho\sigma}\Delta R_{\alpha\beta\gamma\delta} + \mathcal O(\Delta R^3),
\end{equation}
where $\lag_0 = \lag(g_{\mu\nu}, -2/L^2 g_{\mu[\rho}g_{\sigma]\nu})$. To count the power of $\rho$ for each term in the above expansion, we choose $(g_{\mu\nu}, R_{\mu\nu}{}^{\rho\sigma})$ as the fundamental variables of the Lagrangian. In this way we have \cite{Padmanabhan:2013xyr} \footnote{Note that this does not mean the metric not appear explicitly in the Lagrangian, since the derivative of the inverse metric may cancel those of the metric and result in \eqref{eq:lagpdrg-is-0}. An example is $\lag = R_{\mu\nu}{}^{\rho\sigma} R_{\alpha\beta}{}^{\mu\nu}R_\rho{}^{\alpha\gamma\delta}R_{\gamma\delta\sigma}{}^\beta$.} 
\begin{equation}\label{eq:lagpdrg-is-0}
    \left[\frac{\partial\lag}{\partial g_{\mu\nu}}\right]_{g_{\mu\nu}, R_{\mu\nu}{}^{\rho\sigma}} = 0.
\end{equation}
The expansion becomes
\begin{equation}
    \lag(g_{\mu\nu}, R_{\mu\nu\rho\sigma}) = \lag_0 + \hat P^{\mu\nu}{}_{\rho\sigma}\Delta R_{\mu\nu}{}^{\rho\sigma} + \frac{1}{2!}\hat P_2^{\mu\nu}{}_{\rho\sigma}{}^{\alpha\beta}{}_{\gamma\delta}\Delta R_{\mu\nu}{}^{\rho\sigma}\Delta R_{\alpha\beta}{}^{\gamma\delta} + \mathcal O(\Delta R^3).
\end{equation}
From \eqref{eq:lagpdrg-is-0} we know that all the $\hat P_n$ does not depend on $\rho$. On the other hand, it follows from \eqref{eq:delta-r-comps} that $\Delta R_{\mu\nu}{}^{\rho\sigma} = \mathcal O(\rho)$, thus the $n$-th power of $\Delta R$ in the expansion is of order $\mathcal O(\rho^n)$. Therefore, to extract the Weyl anomaly in 8d, we need to consider the first five terms in the series. This means we need to deal with all the tensors $\hat P_n$ where $n \leqslant 4$. Since $P$, $P_2$, $P_3$, $P_4$ has 1, 3, 8, 26 tensor structures respectively, the holographic central charges in 8d could involve at most $1 + 3 + 8 + 26 = 38$ theory-dependent parameters. However, as we shall see the next section, not all of them appear in the central charges.

Our next step is applying the power expansion of the metric \eqref{eq:fg-metric-expand} and pick the anomaly term from the Lagrangian. By solving the equations of motion order by order in $\rho$, we can express the subleading boundary metrics $g^{(i)}_{ab}$ where $i \geqslant 1$ in terms of $g_{ab}^{(0)}$. Their forms are also constrained by PBH transformation, where $g_{ab}^{(1)}$ is completely fixed \cite{Imbimbo:1999bj}
\begin{equation}\label{eq:pdrg1-sol}
    g_{ab}^{(1)} = -\frac{L^2}{d - 2}\left[R_{ab} - \frac{1}{2(d - 1)}R g_{ab}^{(0)}\right] = -L^2 P_{ab},
\end{equation}
where $R_{ab}$, $R$, $P_{ab}$ are the Ricci tensor, Ricci scalar, Schouten tensor of the boundary metric $g_{ab}^{(0)}$. The next-order $g_{ab}^{(2)}$ can be fixed up to two parameters $b_1, b_2$ \cite{Imbimbo:1999bj}
\begin{equation}\label{eq:pdrg2-general-sol}
    g_{ab}^{(2)} = b_1\, C_{cdef} C^{cdef} g_{ab}^{(0)} + b_2\, C_{acde} C_b{}^{cde} - \frac{L^4}{4(d - 4)}B_{ab} + \frac14 P_{ac}P^c{}_b,
\end{equation}
where
\begin{equation}
    B_{ab} = \Box P_{ab} - \nabla^c \nabla_b P_{ac} - C_{cbad} P^{db}
\end{equation}
is the Bach tensor. All the curvature tensors and the covariant derivative $\nabla_a$ are defined by the boundary metric $g_{ab}^{(0)}$. For the 4d and 6d cases, the Weyl anomaly parts of the expanded action include the subleading boundary metrics up to $g_{ab}^{(3)}$, but after substituting the solution of $g_{ab}^{(1)}$ and dropping total derivative terms, they all drop out. We thus do not need to solve their equation of motion. However, in the 8d case although $g_{ab}^{(3)}$ and $g_{ab}^{(4)}$ do not contribute to the action, $g_{ab}^{(2)}$ no longer does. Thus it's mandatory to study the equation of motion.

The equation of motion of the subleading metrics $g_{ab}^{(1, 2, \cdots)}$ can be obtained by expanding \eqref{eq:ghc-eom}
\begin{align}
    \mathcal E^\mu_\nu & = (\hat P^{\mu\rho\sigma\delta} + \hat P_2^{\mu\rho\sigma\delta\alpha\beta\gamma\kappa}\Delta R_{\alpha\beta\gamma\kappa} + \cdots)(\hat R_{\nu\rho\sigma\delta} + \Delta R_{\nu\rho\sigma\delta}) - \frac12\delta^\mu_\nu(\lag_0 + \hat P^{\rho\sigma\alpha\beta}\Delta R_{\rho\sigma\alpha\beta} + \cdots)\nonumber\\
    & \qquad - 2(\hat P_2^{\mu\rho\sigma}{}_\nu{}^{\alpha\beta\gamma\kappa}\nabla_\rho\nabla_\sigma\Delta R_{\alpha\beta\gamma\kappa} + \cdots).
\end{align}
The leading order term in $\rho$ gives $\lag_0 = -2d k_{1,1}/L^2$, similar to \eqref{eq:lag-vac-sol}, and the subleading $\rho^n$ terms give the equation of motion for $g_{ab}^{(n)}$. An alternative and simpler method to derive the equation of motion is varying the Lagrangian-expanded action $\sqrt{|g|}\lag$ with $\lag$ given by \eqref{eq:lag-r-expand} with respect to $g_{ab}(\rho)$ and then apply the metric expansion \eqref{eq:fg-metric-expand}. We apply both methods and cross check each other. We find the solution of $g_{ab}^{(2)}$ indeed has the form \eqref{eq:pdrg2-general-sol}, with $b_1$ and $b_2$ given by \eqref{eq:b1-val} and \eqref{eq:b2-val} in Appendix \ref{sec:ap:results}.

We have now expressed the Weyl anomaly contribution of the action in terms of the boundary curvature fields, plus a total derivative term involving $g^{(3)}_{ab}$
\begin{equation}
    \braket{T_a^a} = \sum_{i = 1}^{92} a_i\mathcal R^{(4)}_i + \nabla_a J^a.
\end{equation}
The central charges can be obtained by recombining the above expression into the form of \eqref{eq:trace-anom-form}. Since both expressions have total derivative contribution, we need to separate total derivative terms and non-total derivative ones and compare the coefficients of non-total derivative terms.

To proceed we need to find all the 8d type B Weyl anomalies $I_i^{(8)}$, which is the main task of the next section.

\section{Weyl anomalies in 8d}
\label{sec:8d-weyl-anoms}
In this section we identify all possible Weyl anomalies in 8d. An algebraic method may be used to find the Weyl anomalies in arbitrary even dimensions \cite{Boulanger:2007st}. Here, however, we choose direct method instead. Let $W[g_{ab}]$ be the effective action of the \ac{cft} where $g_{ab}$ is the background metric. The presence of the Weyl anomaly means the variation of $W[g_{ab}]$ under a conformal transformation $\delta_\sigma g_{ab} = 2\sigma g_{ab}$ is non-zero
\begin{equation}
    \delta_\sigma W[g_{ab}] = \int\dd^d x\sqrt{|g|}\,\sigma \mathcal A.
\end{equation}
The Weyl anomaly is then related to the trace of the energy-momentum tensor
\begin{equation}
    \delta_\sigma W[g_{ab}] = \int\dd^d x\frac{\delta W[g_{ab}]}{\delta g_{ab}}2\sigma g_{ab} = \int\dd^d x\sqrt{|g|}\sigma\braket{T_a^a} \implies \braket{T_a^a} = \mathcal A + (\text{total derivatives}).
\end{equation}
The form of $\mathcal A$ is constrained by certain integrability conditions, namely the Wess-Zumino consistency condition \cite{Wess:1971yu}
\begin{equation}
    [\delta_{\sigma_1}, \delta_{\sigma_2}]W[g_{ab}] = 0.
\end{equation}
In 8d, due to dimensional consideration, $\mathcal A$ should be quartic in curvature, \textit{i.e.}, it can be expressed as a linear combination of the 92 elements in $\mathcal R^{(4)}$ as defined in \eqref{eq:r4-list}
\begin{equation}
    \mathcal A = \sum_{i = 1}^{92} a_i \mathcal R_i^{(4)}.
\end{equation}
The Wess-Zumino condition then reads
\begin{equation}
    [\delta_{\sigma_1}, \delta_{\sigma_2}]W[g_{ab}] = \int\dd^d x \sqrt{|g|}\,\sigma_2 \sum_{i = 1}^{228}\sum_{j = 1}^{92} \mathcal H_i(\sigma_1) M_{ij} a_j,
\end{equation}
where $\mathcal H(\sigma_1)$ is the basis of all the scalars built from the curvature and $\sigma_1$ and their derivatives. We find $\mathcal H(\sigma_1)$ has 228 elements so the dimension of the matrix $M$ is 228 by 92. The complete list of $\mathcal H(\sigma_1)$ is too lengthy and irrelevant so we omit it. Weyl anomalies in 8d is then given by the kernel of $M$, whose dimension is 43. So we get 43 linearly independent Weyl anomalies.

As mentioned earlier, there are trivial anomalies in these 43 Weyl anomalies. They correspond to the third term in \eqref{eq:trace-anom-form}. They are expressible as variations of local terms, thus we can identify them by varying the most general quartic action
\begin{equation}
    \delta_\sigma\int\dd^d x\sqrt{|g|}\sum_{i=1}^{92} a_i\mathcal R^{(4)}_i = \int\dd^d x\sqrt{|g|}\sigma \sum_{i,j=1}^{92}\mathcal R_i^{(4)} B_{ij}a_j.
\end{equation}
We find the rank of the matrix $B$ is 32, so there are 32 trivial anomalies. A basis for them is obtained by applying Gauss elimination on $B_{ij}$. As expected, all of them can be expressed as total derivatives. In this way, we get $43-32=11$ non-trivial Weyl anomalies in 8d.

Now we need to find a proper set of basis for these 11 Weyl anomalies. The first basis is the type A anomaly, which is the 8d Euler density
\begin{align}
    E^{(8)} & = 8! R_{a_1b_1}{}^{[a_1b_1} R_{a_2b_2}{}^{a_2b_2} R_{a_3b_3}{}^{a_3b_3} R_{a_4b_4}{}^{a_4b_4]}\nonumber\\
    & = \mathcal{R}_1^{(4)}-24 \mathcal{R}_2^{(4)}+64 \mathcal{R}_3^{(4)}+48 \mathcal{R}_4^{(4)}-96 \mathcal{R}_5^{(4)}+96 \mathcal{R}_6^{(4)}-384 \mathcal{R}_7^{(4)}+6 \mathcal{R}_8^{(4)}-96 \mathcal{R}_9^{(4)}\nonumber\\
    & \qquad -24 \mathcal{R}_{10}^{(4)}+192 \mathcal{R}_{11}^{(4)}+96 \mathcal{R}_{12}^{(4)}-192 \mathcal{R}_{13}^{(4)}+192 \mathcal{R}_{14}^{(4)}+16 \mathcal{R}_{15}^{(4)}-32 \mathcal{R}_{16}^{(4)}+192 \mathcal{R}_{17}^{(4)}\nonumber\\
    & \qquad -192 \mathcal{R}_{18}^{(4)}+384 \mathcal{R}_{19}^{(4)}+3 \mathcal{R}_{20}^{(4)}-48 \mathcal{R}_{21}^{(4)}+6 \mathcal{R}_{22}^{(4)}+48 \mathcal{R}_{23}^{(4)}-96 \mathcal{R}_{24}^{(4)}+48 \mathcal{R}_{25}^{(4)}\nonumber\\
    & \qquad -96 \mathcal{R}_{26}^{(4)}.
\end{align}
The rest 10 Weyl anomalies are type B ones, \textit{i.e.}, those associated with the Weyl invariants. Seven of them can be identified as quartic scalars constructed solely from non-differentiated Weyl tensors
\begingroup\allowdisplaybreaks
\begin{align}
    I^{(8)}_1 & = (C^{abcd}C_{abcd})^2,\nonumber\\
    I^{(8)}_2 & = C^{abcd}C_{abc}{}^{e}C_{d}{}^{fgh}C_{efgh},\nonumber\\
    I^{(8)}_3 & = C^{abcd}C_{ab}{}^{ef}C_{cd}{}^{gh}C_{efgh},\nonumber\\
    I^{(8)}_4 & = C^{abcd}C_{ab}{}^{ef}C_{ce}{}^{gh}C_{dfgh},\nonumber\\
    I^{(8)}_5 & = C^{abcd}C_{ab}{}^{ef}C_{c}{}^{g}{}_{e}{}^{h}C_{dgfh},\nonumber\\
    I^{(8)}_6 & = C^{abcd}C_{a}{}^{e}{}_{c}{}^{f}C_{b}{}^{g}{}_{d}{}^{h}C_{egfh},\nonumber\\
    I^{(8)}_7 & = C^{abcd}C_{a}{}^{e}{}_{c}{}^{f}C_{b}{}^{g}{}_{e}{}^{h}C_{dgfh}.\label{eq:first7-I8}
\end{align}\endgroup
We are left with the remaining 3 Weyl anomalies containing the covariant derivative. To find a proper Weyl invariant basis for them we need to find all the Weyl invariants in 8d. Although this problem is already addressed in previous work \cite{Boulanger:2004zf}, we rederive the result in a more explicit way. This is done by solving the following linear system
\begin{equation}
    0 = \delta_\sigma\left[\sqrt{|g|}\sum_{i=1}^{92} a_i\mathcal R^{(4)}_i\right] = \sqrt{|g|}\sum_{i=1}^{92} a_i (\delta_\sigma\mathcal R_i^{(4)} + 8\sigma\mathcal R_i^{(4)}) = \sqrt{|g|}\sum_{i=1}^{151}\sum_{j=1}^{92}\mathcal H_i'(\sigma) K_{ij} a_j.
\end{equation}
Here $\mathcal H_i'(\sigma)$ is a subset of $\mathcal H_i(\sigma)$. The linear space of the Weyl anomalies is then given by the kernel of $K_{ij}$, which has dimension 12. Apart from the 7 scalars $I^{(8)}_{1,\cdots, 7}$ we have additionally 5 Weyl invariants with covariant derivatives. This means there are two linearly independent trivial anomalies that are also Weyl invariant. We call them $I_{11}^{(8)}$ and $I_{12}^{(8)}$ respectively, and they are given by
\begin{align}
    I_{11}^{(8)} & = \nabla_a\Big(4\nabla_{f}P_{de}C^{abcd}C_{b}{}^{e}{}_{c}{}^{f}-4\nabla_{e}P_{df}C^{abcd}C_{b}{}^{e}{}_{c}{}^{f}-\nabla_{f}P^{a}{}_{e}C^{bcde}C_{bcd}{}^{f}\nonumber\\
    & \qquad +\nabla^{a}P_{ef}C^{bcde}C_{bcd}{}^{f}+\nabla^{f}C^{bcde}C^{ag}{}_{bc}C_{dfeg}\Big),\nonumber\\
    I_{12}^{(8)} & = \nabla_a\Big(-4\nabla_{f}P_{de}C^{abcd}C_{b}{}^{e}{}_{c}{}^{f}+2\nabla_{e}P_{df}C^{abcd}C_{b}{}^{e}{}_{c}{}^{f}+2\nabla_{d}P_{ef}C^{abcd}C_{b}{}^{e}{}_{c}{}^{f}\nonumber\\
    & \qquad +\nabla^{f}C^{bcde}C^{a}{}_{bd}{}^{g}C_{cgef}-\nabla^{f}C^{bcde}C^{ag}{}_{bc}C_{dfeg}\Big).\label{eq:i1112}
\end{align}
For the remaining three non-trivial Weyl invariants $I_{8,9,10}^{(8)}$, we have no further criteria at this stage to classify them. We shall arbitrarily pick a basis to continue the calculation. As we shall see in the next section, we can recombine them so that their associated central charges becomes proportional to $C_T$, $C_T t_2$, $C_T t_4$ respectively.

\section{Result and discussion}
\label{sec:result}

\subsection{Central charges in 8d}
We are now ready to present our results. The $a$-charge is given by
\begin{equation}
    a = \frac{L^7\pi^4}{36}k_{1,1},
\end{equation}
which is proportional to the holographic entanglement entropy across a spherical region, as expected \cite{Casini:2011kv,Dowker:2010yj,Li:2021jfh}
\begin{equation}
    S \propto -P^{\mu\nu\rho\sigma}\epsilon_{\mu\nu}\epsilon_{\rho\sigma} = 2k_{1,1},
\end{equation}
where the anti-symmetric tensor $\epsilon_{\mu\nu}$ satisfies $\epsilon_{\mu\nu}\epsilon^{\mu\nu} = -2$. The first 7 $c$-charges associated with non-differentiated Weyl invariants \eqref{eq:first7-I8} are given by
\begingroup\allowdisplaybreaks
\begin{align}
    c_1 & = \frac{1}{12} \pi ^4 L^7 k_{1,1}+\frac{1024}{L^3} \pi ^4 \Big[16 b_1^2 \left(100 k_{2,1}+22 k_{2,2}+9 k_{2,3}\right)+4 b_2 b_1 \left(100 k_{2,1}+22 k_{2,2}+9 k_{2,3}\right)\nonumber\\
    & \qquad +b_2^2 \left(25 k_{2,1}+5 k_{2,2}+k_{2,3}\right)\Big]+256 \pi ^4 L \left(b_1 k_{2,3}+\frac{1}{4} b_2 k_{2,3}+k_{4,20}\right)+\frac{256}{L} \pi ^4 \Big[56 b_1^2 k_{1,1}\nonumber\\
    & \qquad +b_1 \left(14 b_2 k_{1,1}+80 k_{3,5}+12 k_{3,6}+12 k_{3,7}-3 k_{3,8}\right)+b_2 \left(b_2 k_{1,1}+10 k_{3,5}+k_{3,6}\right)\Big],\nonumber\\
    c_2 & = -6 \pi ^4 L^7 k_{1,1}+\frac{256 }{L}\pi ^4 b_2 \left(-b_2 k_{1,1}+4 k_{3,6}+12 k_{3,7}-3 k_{3,8}\right)+\frac{2048}{L^3}\pi ^4 b_2^2 \left(2 k_{2,2}+5 k_{2,3}\right)\nonumber\\
    & \qquad +256 \pi ^4 L \left(k_{4,21}-b_2 k_{2,3}\right)-32 \pi ^4 L^5 k_{2,3}+80 \pi ^4 L^3 \left(4 k_{3,7}-k_{3,8}\right),\nonumber\\
    c_3 & = \frac{17}{18} \pi ^4 L^7 k_{1,1}+8 \pi ^4 L^5 k_{2,3}-64 \pi ^4 L^3 k_{3,7}+256 \pi ^4 L k_{4,22},\nonumber\\
    c_4 & = \frac{52}{9} \pi ^4 L^7 k_{1,1}+16 \pi ^4 L^5 k_{2,3}+256 \pi ^4 L^3 \left(\frac{7 k_{3,8}}{16}-\frac{5 k_{3,7}}{4}\right)+256 \pi ^4 L k_{4,23},\nonumber\\
    c_5 & = \frac{100}{9} \pi ^4 L^7 k_{1,1}+192 \pi ^4 L^5 k_{2,3}-32 \pi ^4 L^3 \left(28 k_{3,7}+k_{3,8}\right)+256 \pi ^4 L k_{4,24},\nonumber\\
    c_6 & = \frac{20}{9} \pi ^4 L^7 k_{1,1}+64 \pi ^4 L^5 k_{2,3}-128 \pi ^4 L^3 k_{3,8}+256 \pi ^4 L k_{4,25},\nonumber\\
    c_7 & = -\frac{32}{9} \pi ^4 L^7 k_{1,1}-64 \pi ^4 L^5 k_{2,3}+128 \pi ^4 L^3 k_{3,8}+256 \pi ^4 L k_{4,26}.
\end{align}\endgroup
The last 3 $c$-charges are linear combinations of the four parameters $k_{1,1}$, $k_{2,3}$, $k_{3,7}$, $k_{3,8}$. In this way, we find that the central charges only involve the following 15 theory-dependent parameters
\begin{equation}
\{k_{1,1},k_{2,1},k_{2,2},k_{2,3},k_{3,5},k_{3,6},k_{3,7},k_{3,8},k_{4,20},k_{4,21},k_{4,22},k_{4,23},
k_{4,24},k_{4,25},k_{4,26}\}.
\end{equation}
Interestingly, we find that the last 3 $c$-charges are all linear combinations of $(C_T, C_T t_2, C_T t_4)$, where $t_2$ and $t_4$ are energy flux parameters related to the three-point function of energy-momentum tensor. Details of the calculation of $t_2$ and $t_4$ can be found in appendix \ref{sec:ap:t3pt}. It's expected that $C_T$ is a linear combination of the last 3 $c$-charges, but it's remarkable that the same is true for $C_T t_2$ and $C_T t_4$. This allows us to recombine the last 3 Weyl invariants so that their $c$-charges become
\begin{align}
    c_8 & = \frac{56}{9} \pi ^4 L^5 k_{2,3}-\frac{1568}{3} \pi ^4 L^3 k_{3,7}-280 \pi ^4 L^3 k_{3,8} = c_{10}t_2,\nonumber\\
    c_9 & = 1680 \pi^4 L^3(2 k_{3,7}+ k_{3,8}) = c_{10}t_4,\nonumber\\
    c_{10} & = \frac{L^7\pi^4}{36}(k_{1,1} + 24L^{-2}k_{2,3}) = \frac{\pi^8}{622080}\left.C_T\right|_{d=8}.
\end{align}
We have chosen the coefficient of $c_{10}$ so that it's proportional to $C_T$ and becomes equal to the $a$-charge when restricted to Einstein gravity (\textit{i.e.}, setting $k_{1,1} = 1$ and all other $k$'s vanish). In other words, $c_{10}$ is the 8d analogue of the $c$-charge in 4d.

The corresponding Weyl invariants $I_{8,9,10}^{(8)}$ are given in \eqref{eq:i8-vec}, \eqref{eq:i9-vec}, \eqref{eq:i10-vec}. As we have pointed it out in the abstract, the Weyl invariants are linear combinations of scalar polynomials of the Riemann tensor and covariant derivatives. We use the notation $W_{(n)}$ to denote the Weyl invariants with a minimum number $n$ Riemman tensor in at least one scalar polynomial. Thus the Weyl invariants of \eqref{eq:first7-I8} belong to the $W_{(4)}$ class. We find that $I_{10}^{(8)}$ is $W_{(2)}$ and so contains at most four covariant derivatives, while $I_8^{(8)}$ and $I_9^{(8)}$ are $W_{(3)}$ and contain at most two covariant derivatives. We may then conjecture a general connection between the $c$-charges and energy-momentum tensor correlators in arbitrary even dimensions: The corresponding $c$-charges of $W_{(n)}$ are related to the $n$-point function of energy-momentum tensor. The $n$-point function thus provides a way to further classify the $W_{(n)}$ invariants. Since the two-point function has only one parameter $C_T$, we therefore expect, if the conjecture holds, there's only one $W_{(2)}$ type B Weyl anomaly in arbitrary even dimensions, and the corresponding $c$-charge is proportional to $C_T$. For the 6d case the three $c$-charges are linear combinations of the three-point function parameters $(C_T, C_Tt_2, C_Tt_4)$ (see appendix \ref{eq:ap:6d-weyl-anom}), with $W_{(2)}$ corresponding to $C_T$ and the other two $W_{(3)}$ invariants corresponding to $C_T t_2$ and $C_T t_4$. For the 8d case, we have seen that the conjecture is true for the $W_{(2)}$ and $W_{(3)}$ Weyl invariants, while for $W_{(4)}$ Weyl invariants $I_{1,\cdots, 7}^{(8)}$ the corresponding $c$-charges should be related to the four-point function of energy-momentum tensor. However, conformal invariance does not uniquely fix the form of four-point functions of energy-momentum tensor. Instead, it's determined by a set of scalar functions. The connection between quartic Weyl invariant $c$-charges and four-point functions is not obvious. From the holographic point of view, on the other hand, our conjecture is reasonable since the $W_{(n)}$ Weyl invariants are contributed by at least the tensor $P_n$, which encodes the $n$-th order metric perturbation and thus $n$-point function of energy-momentum tensor. At \ac{cft} side this seems highly non-trivial, and it would be interesting the rederive the same results from \ac{cft}.

\subsection{Einstein gravity and obstruction tensor}
In this subsection we consider the special case of Einstein gravity, which is done by setting $k_{1,1} = 1$ and all other parameters vanish. It was shown that for Einstein gravity the Weyl anomaly could be built from the Schouten tensor and \textit{(extended) obstruction tensor} $\Omega^{(n)}_{ab}$ where $n = 1, 2, \cdots$. The first two obstruction tensors are defined by \cite{graham2008extendedobstructiontensorsrenormalized,Jia:2021hgy}
\begin{align}
    \Omega^{(1)}_{ab} & = -\frac{1}{d - 4}B_{ab},\nonumber\\
    \Omega^{(2)}_{ab} & = \Box B_{ab} - 2C_{dabc}B^{cd} - 4P B_{ab} + 2P_{c(b}B^c{}_{a)} - 2B^c{}_{(a}P_{b)c} + 2(d - 4)\Big[\nabla^d C_{dc(a}P^c{}_{b)}\nonumber\\
    & \qquad - P^{dc}\nabla_{(a}C_{b)cd} + 2P^{dc}\nabla_c C_{(ab)d} + \nabla_c P^{dc}C_{(ab)c} - C^c{}_a{}^dC_{dbc} + \nabla^dP^c{}_{(a}C_{b)cd}\nonumber\\
    & \qquad - C_{c(ab)d}P^c{}_e P^{ec}\Big],
\end{align}
where $P = P_a^a$, and $C_{abc} = 2\nabla_{[c}P_{]ba}$ is the Cotton tensor.

After restricting to Einstein gravity, we find the Weyl anomaly takes the following simple form
\begin{align}
    L^{-7}\braket{T_a^a} & = -\frac{1}{48}\tr(P)^4 + \frac{1}{24}\tr(\Omega^{(1)2}) + \frac{1}{6}\tr(P)\tr(P\Omega^{(1)}) + \frac{1}{24}\tr(P\Omega^{(2)}) + \frac18\tr(P)^2\tr(P^2)\nonumber\\
    & - \frac{1}{16}\tr(P^2)^2 - \frac16\tr(P^2\Omega^{(1)}) - \frac16\tr(P)\tr(P^3) + \frac18 \tr(P^4) + (\text{total derivatives}).
\end{align}
The form of the above result coincides with ref.~\cite{Jia:2021hgy}, obtained using Weyl connection formalism, so they are consistent with each other after going back to the Weyl gauge. Unfortunately, the above simplification does not happen in general higher curvature case. More precisely, the holographic Weyl anomaly from general higher curvature gravity involves 15 theory-dependent constants and 11 anomaly structures. Hence it cannot be expressed only in terms of the following 9 scalar quantities constructed from the Schouten tensor and obstruction tensors
\begin{gather}
    \{\tr(P)^4, \tr(\Omega^{(1)2}), \tr(P)\tr(P\Omega^{(1)}), \tr(P\Omega^{(2)}), \tr(P)^2\tr(P^2),\nonumber\\
    \tr(P^2)^2, \tr(P^2\Omega^{(1)}), \tr(P)\tr(P^3), \tr(P^4)\}.
\end{gather}
This means the general structure of Weyl anomaly proposed by ref.~\cite{Jia:2021hgy} may only holds for Einstein gravity. This is another example of higher curvature gravity bringing new structures to the dual \ac{cft}.

\section{Conclusion}
\label{sec:conclusion}

We calculated the holographic Weyl anomaly of 8d CFT and identified the central charges from the general higher curvature gravity. The general forms of the derivative tensors $P_n$ we introduced allow us to apply the calculation for general higher curvature gravity directly instead of considering a specific Lagrangian. The result is expressed in terms of a set of 15 theory-dependent parameters $k_{n,i}$. Owing to the existence of large number of central charges, we could not adopt the simpler reduced \ac{fg} expansion, which was rather rather effective for 4d and 6d cases. Instead, we adopted the conventional \ac{fg} expansion approach. As a prerequisite of our work, we elaborated the general structures of 8d Weyl anomaly and found 11 non-trivial ones, 10 of which are type B Weyl invariant ones. Since there are a total 12 Weyl invariants in 8d, we have 2 Weyl invariant trivial anomalies that are total derivatives. It may be interesting to further study the properties of these Weyl invariant trivial anomalies, but in this paper, they were simply discarded.

Restricting our results to Einstein gravity we found the Weyl anomaly agreed with that of ref.~\cite{Jia:2021hgy}, which is expressed more compactly in terms of Schouten and obstruction tensors. However, for the general higher curvature case, the Weyl anomaly has more structures and hence it cannot be expressed as scalar quantities built only from Schouten and obstruction tensors.

Among the 10 non-trivial type B anomalies, three of them contain covariant derivatives and, remarkably, the corresponding $c$-charges are linear combinations of $(C_T, C_T t_2, C_T t_4)$. We then recombined the last three Weyl invariants so that their $c$-charges become $c_{10}t_2, c_{10} t_4$ and $c_{10}$ respectively, where $c_{10} \propto C_T$. We found that $I_{10}^{(8)}$ was $W_{(2)}$, while $I_8^{(8)}$ and $I_9^{(8)}$ were $W_{(3)}$. Together with the analogous results from 4d and 6d Weyl anomalies, we make a conjecture that the $c$-charges of $W_{(n)}$ invariants are related to the energy-momentum $n$-point function. A corollary of this conjecture is that there should be only one $W_{(2)}$ type B Weyl anomaly in arbitrary even dimensions and the corresponding $c$-charge is proportional to $C_T$. It is of great interest to further test this conjecture, \textit{e.g.} investigating the connection between the corresponding $c$-charges of the quartic Weyl invariants $c_{1,\cdots,7}$ and energy-momentum four-point function.

\section{Acknowledgement}
We thank Wei-Zhen Jia for helpful discussions. This work is supported in part by the National Natural Science Foundation of China (NSFC) grants No.~11935009 and No.~12375052.

\appendix
\section{List of results}
\label{sec:ap:results}
In this appendix section, we present the part of our results that are too lengthy to put in the main text.

\subsection{Riemann scalars}
\label{sec:ap:reimann-scalars}

In this paper, particularly in section \ref{sec:method}, we introduced scalar quantities $\mathcal R^{(n)}$ built from linear combinations of polynomial invariants of Riemann tensor and covariant derivatives, where $n$ denotes that each polynomial involves a total of $2n$ derivatives, with Riemann tensor and a covariant derivative contributing 2 and 1 derivatives respectively. The bases for $n=1,2,3$ are simpler and given in \eqref{eq:r123-list}. The basis for $\mathcal R^{(4)}$ is more lengthy, and we have totally 26 non-differentiated terms, plus 66 differentiated terms
\begingroup\allowdisplaybreaks
\begin{align}
    \mathcal R^{(4)} & = \{R^4,R^2R^{\mu\nu}R_{\mu\nu},RR^{\mu\nu}R_{\mu}{}^{\rho}R_{\nu\rho},(R^{\mu\nu}R_{\mu\nu})^2,R^{\mu\nu}R_{\mu}{}^{\rho}R_{\nu}{}^{\sigma}R_{\rho\sigma},RR_{\mu\rho}R_{\nu\sigma}R^{\mu\nu\rho\sigma},\nonumber\\
    & \qquad R_{\mu\rho}R_{\nu}{}^{\alpha}R_{\sigma\alpha}R^{\mu\nu\rho\sigma},R^2R^{\mu\nu\rho\sigma}R_{\mu\nu\rho\sigma},RR_{\sigma\alpha}R^{\mu\nu\rho\sigma}R_{\mu\nu\rho}{}^{\alpha},(R^{\mu\nu}R_{\mu\nu})(R^{\mu\nu\rho\sigma}R_{\mu\nu\rho\sigma}),\nonumber\\
    & \qquad R_{\alpha\beta}R_{\sigma}{}^{\beta}R^{\mu\nu\rho\sigma}R_{\mu\nu\rho}{}^{\alpha},R_{\rho\alpha}R_{\sigma\beta}R^{\mu\nu\rho\sigma}R_{\mu\nu}{}^{\alpha\beta},R_{\alpha\beta}R_{\nu\sigma}R^{\mu\nu\rho\sigma}R_{\mu}{}^{\alpha}{}_{\rho}{}^{\beta},\nonumber\\
    & \qquad R_{\nu\alpha}R_{\sigma\beta}R^{\mu\nu\rho\sigma}R_{\mu}{}^{\alpha}{}_{\rho}{}^{\beta},RR^{\mu\nu\rho\sigma}R_{\mu\nu}{}^{\alpha\beta}R_{\rho\sigma\alpha\beta},RR^{\mu\nu\rho\sigma}R_{\mu}{}^{\alpha}{}_{\rho}{}^{\beta}R_{\nu\alpha\sigma\beta},\nonumber\\
    & \qquad R_{\beta\gamma}R^{\mu\nu\rho\sigma}R_{\mu\nu\rho}{}^{\alpha}R_{\sigma}{}^{\beta}{}_{\alpha}{}^{\gamma},R_{\beta\gamma}R^{\mu\nu\rho\sigma}R_{\mu\nu}{}^{\alpha\beta}R_{\rho\sigma\alpha}{}^{\gamma},R_{\beta\gamma}R^{\mu\nu\rho\sigma}R_{\mu}{}^{\alpha}{}_{\rho}{}^{\beta}R_{\nu\alpha\sigma}{}^{\gamma},\nonumber\\
    & \qquad (R^{\mu\nu\rho\sigma}R_{\mu\nu\rho\sigma})^2,R^{\mu\nu\rho\sigma}R_{\alpha\beta\gamma\delta}R_{\mu\nu\rho}{}^{\alpha}R_{\sigma}{}^{\beta\gamma\delta},R^{\mu\nu\rho\sigma}R_{\alpha\beta\gamma\delta}R_{\mu\nu}{}^{\alpha\beta}R_{\rho\sigma}{}^{\gamma\delta},\nonumber\\
    & \qquad R^{\mu\nu\rho\sigma}R_{\mu\nu}{}^{\alpha\beta}R_{\rho\alpha}{}^{\gamma\delta}R_{\sigma\beta\gamma\delta},R^{\mu\nu\rho\sigma}R_{\mu\nu}{}^{\alpha\beta}R_{\rho}{}^{\gamma}{}_{\alpha}{}^{\delta}R_{\sigma\gamma\beta\delta},R^{\mu\nu\rho\sigma}R_{\alpha\gamma\beta\delta}R_{\mu}{}^{\alpha}{}_{\rho}{}^{\beta}R_{\nu}{}^{\gamma}{}_{\sigma}{}^{\delta},\nonumber\\
    & \qquad R^{\mu\nu\rho\sigma}R_{\mu}{}^{\alpha}{}_{\rho}{}^{\beta}R_{\nu}{}^{\gamma}{}_{\alpha}{}^{\delta}R_{\sigma\gamma\beta\delta},R\nabla^{\mu}R\nabla_{\mu}R,\nabla_{\mu}R\nabla_{\nu}RR^{\mu\nu},\nabla_{\rho}R\nabla^{\rho}R^{\mu\nu}R_{\mu\nu},\nonumber\\
    & \qquad \nabla_{\nu}R\nabla^{\rho}R^{\mu\nu}R_{\mu\rho},\nabla_{\sigma}R\nabla_{\nu}R_{\mu\rho}R^{\mu\nu\rho\sigma},R\nabla^{\rho}R^{\mu\nu}\nabla_{\rho}R_{\mu\nu},R\nabla^{\rho}R^{\mu\nu}\nabla_{\nu}R_{\mu\rho},\nonumber\\
    & \qquad \nabla^{\rho}R^{\mu\nu}\nabla_{\rho}R_{\mu}{}^{\sigma}R_{\nu\sigma},\nabla^{\rho}R^{\mu\nu}\nabla_{\mu}R_{\rho}{}^{\sigma}R_{\nu\sigma},\nabla^{\rho}R^{\mu\nu}\nabla^{\sigma}R_{\mu\rho}R_{\nu\sigma},\nabla^{\rho}R^{\mu\nu}\nabla^{\sigma}R_{\mu\nu}R_{\rho\sigma},\nonumber\\
    & \qquad \nabla^{\alpha}R_{\mu\rho}\nabla_{\alpha}R_{\nu\sigma}R^{\mu\nu\rho\sigma},\nabla^{\alpha}R_{\mu\rho}\nabla_{\sigma}R_{\nu\alpha}R^{\mu\nu\rho\sigma},\nabla_{\nu}R_{\mu}{}^{\alpha}\nabla_{\sigma}R_{\rho\alpha}R^{\mu\nu\rho\sigma},\nonumber\\
    & \qquad \nabla_{\rho}R_{\mu}{}^{\alpha}\nabla_{\sigma}R_{\nu\alpha}R^{\mu\nu\rho\sigma},\nabla_{\alpha}R\nabla^{\alpha}R^{\mu\nu\rho\sigma}R_{\mu\nu\rho\sigma},\nabla_{\alpha}R_{\mu\rho}\nabla^{\alpha}R^{\mu\nu\rho\sigma}R_{\nu\sigma},\nonumber\\
    & \qquad \nabla_{\rho}R_{\mu\alpha}\nabla^{\alpha}R^{\mu\nu\rho\sigma}R_{\nu\sigma},\nabla_{\alpha}R_{\sigma\beta}\nabla^{\alpha}R^{\mu\nu\rho\sigma}R_{\mu\rho\nu}{}^{\beta},\nabla_{\sigma}R_{\alpha\beta}\nabla^{\alpha}R^{\mu\nu\rho\sigma}R_{\mu\rho\nu}{}^{\beta},\nonumber\\
    & \qquad \nabla_{\beta}R_{\sigma\alpha}\nabla^{\alpha}R^{\mu\nu\rho\sigma}R_{\mu\rho\nu}{}^{\beta},R\nabla^{\alpha}R^{\mu\nu\rho\sigma}\nabla_{\alpha}R_{\mu\nu\rho\sigma},\nabla^{\alpha}R^{\mu\nu\rho\sigma}\nabla_{\alpha}R_{\mu\nu\rho}{}^{\beta}R_{\sigma\beta},\nonumber\\
    & \qquad \nabla^{\alpha}R^{\mu\nu\rho\sigma}\nabla_{\sigma}R_{\mu\alpha\rho}{}^{\beta}R_{\nu\beta},\nabla^{\alpha}R^{\mu\nu\rho\sigma}\nabla_{\alpha}R_{\mu\nu}{}^{\beta\gamma}R_{\rho\sigma\beta\gamma},\nabla^{\alpha}R^{\mu\nu\rho\sigma}\nabla_{\alpha}R_{\mu}{}^{\beta}{}_{\rho}{}^{\gamma}R_{\nu\beta\sigma\gamma},\nonumber\\
    & \qquad \nabla^{\alpha}R^{\mu\nu\rho\sigma}\nabla_{\rho}R_{\mu}{}^{\beta}{}_{\alpha}{}^{\gamma}R_{\nu\beta\sigma\gamma},R^2\Box R,R\nabla^{\nu}\nabla^{\mu}RR_{\mu\nu},\Box R(R^{\mu\nu}R_{\mu\nu}),\nabla^{\nu}\nabla^{\mu}RR_{\mu}{}^{\rho}R_{\nu\rho},\nonumber\\
    & \qquad \nabla_{\rho}\nabla_{\mu}RR_{\nu\sigma}R^{\mu\nu\rho\sigma},\Box R(R^{\mu\nu\rho\sigma}R_{\mu\nu\rho\sigma}),\nabla_{\alpha}\nabla_{\sigma}RR^{\mu\nu\rho\sigma}R_{\mu\nu\rho}{}^{\alpha},(\Box R)^2,\nabla^{\nu}\nabla^{\mu}R\nabla_{\nu}\nabla_{\mu}R,\nonumber\\
    & \qquad R\Box R^{\mu\nu}R_{\mu\nu},\Box R^{\mu\nu}R_{\mu}{}^{\sigma}R_{\nu\sigma},\nabla^{\sigma}\nabla^{\rho}R^{\mu\nu}R_{\mu\nu}R_{\rho\sigma},\nabla^{\sigma}\nabla^{\rho}R^{\mu\nu}R_{\mu\rho}R_{\nu\sigma},\nonumber\\
    & \qquad R\nabla_{\sigma}\nabla_{\nu}R_{\mu\rho}R^{\mu\nu\rho\sigma},\Box R_{\mu\rho}R_{\nu\sigma}R^{\mu\nu\rho\sigma},\nabla_{\nu}\nabla_{\rho}R_{\mu}{}^{\alpha}R_{\sigma\alpha}R^{\mu\nu\rho\sigma},\nabla_{\nu}\nabla^{\alpha}R_{\mu\rho}R_{\sigma\alpha}R^{\mu\nu\rho\sigma},\nonumber\\
    & \qquad \Box R_{\sigma\alpha}R^{\mu\nu\rho\sigma}R_{\mu\nu\rho}{}^{\alpha},\nabla_{\beta}\nabla_{\sigma}R_{\rho\alpha}R^{\mu\nu\rho\sigma}R_{\mu\nu}{}^{\alpha\beta},\nabla_{\beta}\nabla_{\alpha}R_{\nu\sigma}R^{\mu\nu\rho\sigma}R_{\mu}{}^{\alpha}{}_{\rho}{}^{\beta},\nonumber\\
    & \qquad \nabla_{\beta}\nabla_{\sigma}R_{\nu\alpha}R^{\mu\nu\rho\sigma}R_{\mu}{}^{\alpha}{}_{\rho}{}^{\beta},\nabla_{\nu}\nabla_{\mu}R\Box R^{\mu\nu},\Box R^{\mu\nu}\Box R_{\mu\nu},\nabla^{\sigma}\nabla^{\rho}R^{\mu\nu}\nabla_{\sigma}\nabla_{\rho}R_{\mu\nu},\nonumber\\
    & \qquad \nabla^{\sigma}\nabla^{\rho}R^{\mu\nu}\nabla_{\sigma}\nabla_{\nu}R_{\mu\rho},\nabla^{\sigma}\nabla^{\rho}R^{\mu\nu}\nabla_{\nu}\nabla_{\mu}R_{\rho\sigma},\nabla^{\beta}\nabla^{\alpha}R^{\mu\nu\rho\sigma}R_{\nu\sigma}R_{\mu\alpha\rho\beta},\nonumber\\
    & \qquad \nabla^{\beta}\nabla^{\alpha}R^{\mu\nu\rho\sigma}R_{\mu\rho\nu}{}^{\gamma}R_{\sigma\beta\alpha\gamma},\nabla^{\beta}\nabla^{\alpha}R^{\mu\nu\rho\sigma}\nabla_{\beta}\nabla_{\alpha}R_{\mu\nu\rho\sigma},\nabla_{\nu}R\nabla_{\mu}\nabla^{\nu}\nabla^{\mu}R,\nonumber\\
    & \qquad \nabla^{\rho}\nabla^{\nu}\nabla^{\mu}R\nabla_{\rho}R_{\mu\nu},\nabla_{\rho}R_{\mu\nu}\Box \nabla^{\rho}R^{\mu\nu},\nabla_{\nu}R_{\mu\rho}\Box \nabla^{\rho}R^{\mu\nu},\nabla_{\alpha}\nabla_{\sigma}\nabla_{\nu}R_{\mu\rho}\nabla^{\alpha}R^{\mu\nu\rho\sigma},\nonumber\\
    & \qquad R\Box^2R,\Box \nabla^{\nu}\nabla^{\mu}RR_{\mu\nu},\Box^2R^{\mu\nu}R_{\mu\nu},\Box \nabla_{\sigma}\nabla_{\nu}R_{\mu\rho}R^{\mu\nu\rho\sigma},\Box^3R\}.\label{eq:r4-list}
\end{align}
\endgroup
The redundancies due to the Bianchi identities are taken into account.

\subsection{Sub-subleading solution of the metric in FG expansion}
The parameters $b_1$ and $b_2$ in the solution of $g_{ab}^{(2)}$ of \eqref{eq:pdrg2-general-sol} are given by
\begingroup\allowdisplaybreaks
\begin{align}
    b_1 & = \Big[-40 \left(3 d^2-26 d+56\right) L^2 k_{2,3} k_{3,5}+80 \left(d^2-7 d+12\right) L^2 k_{2,1} k_{3,6}\nonumber\\
    & \qquad +4 \left(3 d^2-20 d+32\right) L^2 k_{2,2} k_{3,6}-12 \left(d^2-10 d+24\right) L^2 k_{2,3} k_{3,6}\nonumber\\
    & \qquad +240 \left(d^2-7 d+12\right) L^2 k_{2,1} k_{3,7}+48 \left(d^2-7 d+12\right) L^2 k_{2,2} k_{3,7}\nonumber\\
    & \qquad -60 \left(d^2-7 d+12\right) L^2 k_{2,1} k_{3,8}-12 \left(d^2-7 d+12\right) L^2 k_{2,2} k_{3,8}+4 (10-3 d) L^4 k_{2,3}^2\nonumber\\
    & \qquad -80 (d-3) L^4 k_{2,1} k_{2,3}+4 (16-5 d) L^4 k_{2,2} k_{2,3}+10 (d-4) L^4 k_{1,1} k_{3,5}\nonumber\\
    & \qquad +5 (d-4) L^4 k_{1,1} k_{3,6}+12 (d-4) L^4 k_{1,1} k_{3,7}-3 (d-4) L^4 k_{1,1} k_{3,8}-40 (d-4)^2 L^2 k_{2,2} k_{3,5}\nonumber\\
    & \qquad +48 (d-4) L^2 k_{2,3} k_{3,7}-12 (d-4) L^2 k_{2,3} k_{3,8}-3 L^6 k_{1,1} k_{2,3}\Big] / (M_1M_2),\label{eq:b1-val}\\
    b_2 & = L^2\left[-4 (d-4)k_{3,6}-12 (d-4)k_{3,7}+3 (d-4) k_{3,8}+4 L^2 k_{2,3}\right] / M_2,\label{eq:b2-val}
\end{align}
\endgroup
where
\begin{align}
    M_1 & = 4 d^2 \left(5 k_{2,1}+k_{2,2}\right)+(d-1) L^2 k_{1,1}-4 d \left(15 k_{2,1}+2 k_{2,2}-4 k_{2,3}\right)-8 \left(2 k_{2,2}+7 k_{2,3}\right),\nonumber\\
    M_2 & = 4 (d-4) \left[4 (d-4) k_{2,2}+4 (3 d-14) k_{2,3}-L^2k_{1,1}\right].
\end{align}

\subsection{Weyl invariants in 8d}

There are a total of 12 Weyl invariants in 8d. Seven of these are quartic Weyl invariants, given in \eqref{eq:first7-I8}. Two are total derivatives, given in \eqref{eq:i1112}.  Here, we gave the remaining three invariants, $I_{8,\cdots,12}^{(8)}$ in terms of the Weyl tensor $C_{abcd}$ and Schouten tensor $P_{ab}$, together with covariant derivatives. We also define the trace of the Schouten tensor
\begin{equation}
    P = P_a^a = \frac{R}{2(d - 1)} = \frac{R}{14}.
\end{equation}
We have
\begingroup\allowdisplaybreaks
\begin{align}
    I_8^{(8)} & = -\frac{2376}{7}\nabla^{e}P_{ac}\nabla_{e}P_{bd}C^{abcd}-\frac{2376}{7}\nabla_{c}P_{a}{}^{e}\nabla_{d}P_{be}C^{abcd}+\frac{4752}{7}\nabla^{e}P_{ac}\nabla_{d}P_{be}C^{abcd}\nonumber\\
    & \qquad -\frac{3168}{7}\nabla_{b}P_{a}{}^{e}\nabla_{d}P_{ce}C^{abcd}-\frac{1296}{7}\nabla_{f}\nabla_{e}P_{bd}C^{abcd}C_{a}{}^{e}{}_{c}{}^{f}+\frac{1296}{7}\nabla_{f}\nabla_{d}P_{be}C^{abcd}C_{a}{}^{e}{}_{c}{}^{f}\nonumber\\
    & \qquad -\frac{432}{7}\nabla_{f}\nabla_{d}P_{ce}C^{abcd}C_{ab}{}^{ef}-\frac{2880}{7}\nabla_{f}P_{de}\nabla^{e}C^{abcd}C_{acb}{}^{f}+\frac{1440}{7}\nabla_{e}P_{df}\nabla^{e}C^{abcd}C_{acb}{}^{f}\nonumber\\
    & \qquad +\frac{1440}{7}\nabla_{d}P_{ef}\nabla^{e}C^{abcd}C_{acb}{}^{f}-\frac{1296}{7}P_{fg}C^{abcd}C_{a}{}^{e}{}_{c}{}^{f}C_{bed}{}^{g}-\frac{408}{7}\nabla^{e}C^{abcd}\nabla_{e}C_{a}{}^{f}{}_{c}{}^{g}C_{bfdg}\nonumber\\
    & \qquad +\frac{360}{7}\nabla^{e}C^{abcd}\nabla_{c}C_{a}{}^{f}{}_{e}{}^{g}C_{bfdg}+\frac{108}{7}P_{fg}C^{abcd}C_{ab}{}^{ef}C_{cde}{}^{g}+\frac{24}{7}\nabla^{e}C^{abcd}\nabla_{e}C_{ab}{}^{fg}C_{cdfg},\label{eq:i8-vec}\\
    I_9^{(8)} & = -\frac{11488}{105}\nabla^{e}P_{ac}\nabla_{e}P_{bd}C^{abcd}-\frac{11488}{105}\nabla_{c}P_{a}{}^{e}\nabla_{d}P_{be}C^{abcd}+\frac{22976}{105}\nabla^{e}P_{ac}\nabla_{d}P_{be}C^{abcd}\nonumber\\
    & \qquad -\frac{1024}{7}\nabla_{b}P_{a}{}^{e}\nabla_{d}P_{ce}C^{abcd}-\frac{6208}{105}\nabla_{f}\nabla_{e}P_{bd}C^{abcd}C_{a}{}^{e}{}_{c}{}^{f}+\frac{6208}{105}\nabla_{f}\nabla_{d}P_{be}C^{abcd}C_{a}{}^{e}{}_{c}{}^{f}\nonumber\\
    & \qquad -\frac{1984}{105}\nabla_{f}\nabla_{d}P_{ce}C^{abcd}C_{ab}{}^{ef}-\frac{2816}{21}\nabla_{f}P_{de}\nabla^{e}C^{abcd}C_{acb}{}^{f}+\frac{1408}{21}\nabla_{e}P_{df}\nabla^{e}C^{abcd}C_{acb}{}^{f}\nonumber\\
    & \qquad +\frac{1408}{21}\nabla_{d}P_{ef}\nabla^{e}C^{abcd}C_{acb}{}^{f}-\frac{6208}{105}P_{fg}C^{abcd}C_{a}{}^{e}{}_{c}{}^{f}C_{bed}{}^{g}-\frac{1984}{105}\nabla^{e}C^{abcd}\nabla_{e}C_{a}{}^{f}{}_{c}{}^{g}C_{bfdg}\nonumber\\
    & \qquad +\frac{352}{21}\nabla^{e}C^{abcd}\nabla_{c}C_{a}{}^{f}{}_{e}{}^{g}C_{bfdg}+\frac{16}{3}P_{fg}C^{abcd}C_{ab}{}^{ef}C_{cde}{}^{g}+\frac{8}{7}\nabla^{e}C^{abcd}\nabla_{e}C_{ab}{}^{fg}C_{cdfg},\label{eq:i9-vec}\\
    I_{10}^{(8)} & = -176\nabla^{b}\nabla^{a}P\nabla_{b}\nabla_{a}P+2240\nabla^{d}\nabla^{c}P^{ab}\nabla_{d}\nabla_{c}P_{ab}-3968\nabla^{d}\nabla^{c}P^{ab}\nabla_{d}\nabla_{b}P_{ac}\nonumber\\
    & \qquad +1728\nabla^{d}\nabla^{c}P^{ab}\nabla_{b}\nabla_{a}P_{cd}+8\nabla^{f}\nabla^{e}C^{abcd}\nabla_{f}\nabla_{e}C_{abcd}-2560P\nabla^{c}P^{ab}\nabla_{c}P_{ab}\nonumber\\
    & \qquad +2560P\nabla^{c}P^{ab}\nabla_{b}P_{ac}+640\nabla_{e}\nabla_{d}\nabla_{b}P_{ac}\nabla^{e}C^{abcd}+1280\nabla_{c}P_{ab}\Box \nabla^{c}P^{ab}\nonumber\\
    & \qquad -1280\nabla_{b}P_{ac}\Box \nabla^{c}P^{ab}+352\nabla_{b}\nabla_{a}P\Box P^{ab}-176\Box P^{ab}\Box P_{ab}+1408(P^{ab}P_{ab})^2\nonumber\\
    & \qquad -6272\nabla^{b}\nabla^{a}PP_{a}{}^{c}P_{bc}-2560\nabla^{c}P^{ab}\nabla_{c}P_{a}{}^{d}P_{bd}+8960\nabla^{c}P^{ab}\nabla^{d}P_{ac}P_{bd}\nonumber\\
    & \qquad -1920\nabla^{c}P^{ab}\nabla_{a}P_{c}{}^{d}P_{bd}+1280\nabla_{e}P_{ac}\nabla^{e}C^{abcd}P_{bd}-1280\nabla_{c}P_{ae}\nabla^{e}C^{abcd}P_{bd}\nonumber\\
    & \qquad +6272\Box P^{ab}P_{a}{}^{d}P_{bd}+640\nabla^{e}C^{abcd}\nabla_{d}C_{aec}{}^{f}P_{bf}-4480\nabla^{c}P^{ab}\nabla^{d}P_{ab}P_{cd}\nonumber\\
    & \qquad -11264P^{ab}P_{a}{}^{c}P_{b}{}^{d}P_{cd}+320\nabla^{e}C^{abcd}\nabla_{e}C_{abc}{}^{f}P_{df}-896P\nabla_{d}\nabla_{b}P_{ac}C^{abcd}\nonumber\\
    & \qquad +6944\nabla^{e}P_{ac}\nabla_{e}P_{bd}C^{abcd}+6944\nabla_{c}P_{a}{}^{e}\nabla_{d}P_{be}C^{abcd}-13888\nabla^{e}P_{ac}\nabla_{d}P_{be}C^{abcd}\nonumber\\
    & \qquad +8448\nabla_{b}P_{a}{}^{e}\nabla_{d}P_{ce}C^{abcd}+320\Box \nabla_{d}\nabla_{b}P_{ac}C^{abcd}+1024\nabla_{c}\nabla_{a}PP_{bd}C^{abcd}\nonumber\\
    & \qquad -1024\Box P_{ac}P_{bd}C^{abcd}-8064\nabla_{b}\nabla_{c}P_{a}{}^{e}P_{de}C^{abcd}+1152\nabla_{b}\nabla^{e}P_{ac}P_{de}C^{abcd}\nonumber\\
    & \qquad +11648P_{ac}P_{b}{}^{e}P_{de}C^{abcd}+1472\nabla_{f}\nabla_{e}P_{bd}C^{abcd}C_{a}{}^{e}{}_{c}{}^{f}-1472\nabla_{f}\nabla_{d}P_{be}C^{abcd}C_{a}{}^{e}{}_{c}{}^{f}\nonumber\\
    & \qquad +3648P_{be}P_{df}C^{abcd}C_{a}{}^{e}{}_{c}{}^{f}-1728P_{bd}P_{ef}C^{abcd}C_{a}{}^{e}{}_{c}{}^{f}+320\nabla_{f}\nabla_{d}P_{ce}C^{abcd}C_{ab}{}^{ef}\nonumber\\
    & \qquad +1536P_{ce}P_{df}C^{abcd}C_{ab}{}^{ef}+96\nabla_{e}\nabla_{d}PC^{abcd}C_{abc}{}^{e}+64\Box P_{de}C^{abcd}C_{abc}{}^{e}\nonumber\\
    & \qquad -128PP_{de}C^{abcd}C_{abc}{}^{e}-1344P_{d}{}^{f}P_{ef}C^{abcd}C_{abc}{}^{e}+80\nabla_{e}P\nabla^{e}C^{abcd}C_{abcd}\nonumber\\
    & \qquad +5376\nabla_{f}P_{de}\nabla^{e}C^{abcd}C_{acb}{}^{f}-896\nabla_{e}P_{df}\nabla^{e}C^{abcd}C_{acb}{}^{f}-3200\nabla_{d}P_{ef}\nabla^{e}C^{abcd}C_{acb}{}^{f}\nonumber\\
    & \qquad +384\nabla^{f}\nabla^{e}C^{abcd}P_{bd}C_{aecf}+448P_{fg}C^{abcd}C_{a}{}^{e}{}_{c}{}^{f}C_{bed}{}^{g}+448\nabla^{e}C^{abcd}\nabla_{e}C_{a}{}^{f}{}_{c}{}^{g}C_{bfdg}\nonumber\\
    & \qquad -544\nabla^{e}C^{abcd}\nabla_{c}C_{a}{}^{f}{}_{e}{}^{g}C_{bfdg}-464P_{fg}C^{abcd}C_{ab}{}^{ef}C_{cde}{}^{g}-72\nabla^{e}C^{abcd}\nabla_{e}C_{ab}{}^{fg}C_{cdfg}.\label{eq:i10-vec}
\end{align}\endgroup
It is clear that $I_{8,9}^{(8)}$ and $I_{10}^{(8)}$ belong to $W_{(3)}$ and $W_{(2)}$ respectively.

\section{Energy-momentum tensor three-point function}
\label{sec:ap:t3pt}
Due to conformal invariance the form three-point function of the energy-momentum tensor is fixed up to three parameters $\cal A, B, C$ and can be schematically expressed as \cite{Osborn:1993cr,Erdmenger:1996yc}
\begin{equation}
    \braket{T^{ab}(x)T^{cd}(y)T^{ef}(z)} = \frac{\mathcal A\mathcal I_1^{abcdef} + \mathcal B\mathcal I_2^{abcdef} + \mathcal C\mathcal I_3^{abcdef}}{|x-y|^d|y-z|^d|z-x|^d},
\end{equation}
where $\mathcal I_i^{abcdef}$ are some complicated tensor structures. By Ward identities $\cal(A, B, C)$ are related to $C_T$ through
\begin{equation}\label{eq:ct-from-abc}
    C_T = \frac{\pi^d}{\Gamma(d/2)}\frac{(d - 1)(d + 2)\mathcal A - 2\mathcal B - 4(d + 1)\mathcal C}{d(d + 2)}.
\end{equation}

The parameters $\cal(A, B, C)$ can be calculated holographically using standard method, but the calculation is too challenging for higher curvature gravity. Therefore an indirect method was proposed, by considering the energy flux operator \cite{Myers:2010jv,Hofman:2008ar}
\begin{equation}
    \mathcal E(\vec n) \sim \int\dd y^- T^{--}(y^+ = 0, y^-, y^i),
\end{equation}
which measures the energy flux at future null infinity in the direction $\vec n$. Note that we work in light cone coordinates $(u, y^+, y^-, y^i)$. Then the expectation value of $\mathcal E(\vec n)$ in the state created by $\mathcal O \sim \varepsilon_{ab}T^{ab}$ encodes the three-point function parameters $\cal (A, B, C)$
\begin{equation}\label{eq:energy-flux-general-form}
    \frac{\braket{\mathcal O^\dagger\mathcal E(\vec n)\mathcal O}}{\braket{\mathcal O^\dagger\mathcal O}} \sim 1 + t_2\left(\frac{\varepsilon_{ij}^\ast\varepsilon_{il}n^jn^l}{\varepsilon_{ij}^\ast\varepsilon_{ij}} - \frac{1}{d - 1}\right) + t_4\left(\frac{\left|\varepsilon_{ij}n^in^j\right|^2}{\varepsilon_{ij}^\ast\varepsilon_{ij}} - \frac{2}{d^2 - 1}\right).
\end{equation}
The energy flux parameters $t_2, t_4$ are related to $\cal (A, B, C)$ by \cite{Buchel:2009sk}
\begin{align}
    t_2 & = \frac{(2 (d+1)) \left(\mathcal A (d-2) (d+1) (d+2)+3 \mathcal B d^2-4 \mathcal C (2 d+1) d\right)}{d (\mathcal A (d-1) (d+2)-2 \mathcal B-4 \mathcal C (d+1))},\label{eq:t2-from-abc}\\
    t_4 & = -\frac{(d+1) \left(\mathcal A \left(2 d^2-3 d-3\right) (d+2)+2 \mathcal B (d+2) d^2-4 \mathcal C (d+1) (d+2) d\right)}{d (\mathcal A (d-1) (d+2)-2 \mathcal B-4 \mathcal C (d+1))}.\label{eq:t4-from-abc}
\end{align}

The energy flux operator $\mathcal E(\vec n)$ is sourced by the shock wave solution
\begin{equation}
    \dd s_{\rm shockwave}^2 = \dd s^2_{\rm AdS} + \frac{L^2}{u^2}\delta(y^+)W(u, y^i)(\dd y^+)^2,
\end{equation}
where
\begin{equation}
    \dd s^2_{\rm AdS} = \frac{L^2}{u^2}\left(\dd u^2 - \dd y^+\dd y^- + \dd y^i\dd y^i\right).
\end{equation}
The insertion of $\mathcal O$ is introduced using graviton perturbation. For a higher curvature gravity with concrete Lagrangian, one only need to consider the the second order around the shock wave background. But for our case, since the curvature of the shock wave cannot be expressed into the form $\hat R_{\mu\nu\rho\sigma} = 2\lambda g_{\mu[\rho}g_{\sigma]\nu}$, we have to also regard the shock wave as perturbation and compute third order perturbation around AdS background. After some calculation, we find
\begin{align}
    t_2 & = \frac{(d-1) d \left[4 L^2 k_{2,3}-12 (3 d+4) k_{3,7}-3 (7 d+4) k_{3,8}\right]}{L^2 \left[4 (d-2) k_{2,3}+L^2 k_{1,1}\right]},\\
    t_4 & = \frac{6 (d-1) d (d+1) (d+2) \left(2 k_{3,7}+k_{3,8}\right)}{L^2 \left[4 (d-2) k_{2,3}+L^2 k_{1,1}\right]}.
\end{align}
This result was already derived in ref.~\cite{Sen:2014nfa} and we obtained it independently. The denominators of $t_2$ and $t_4$ are proportional to $C_T$, thus by \eqref{eq:t4-from-abc} is clear that $(C_T, C_T t_2, C_T t_4)$ can be chosen as an alternative basis of the three-point function parameters.

\section{Holographic Weyl anomalies in 6d}
\label{eq:ap:6d-weyl-anom}
In 6d there are three type B Weyl anomalies \cite{Bastianelli:2000rs}
\begin{align}
    I_1^{(6)} & = C^{ab}{}_{cd}C^{cd}{}_{ef}C^{ef}{}_{ab},\\
    I_2^{(6)} & = C_{abcd}C^{aefd}C_e{}^{bc}{}_f,\\
    I_3^{(6)} & = \Box C^{abcd}C_{abcd} - \frac65 R C^{abcd}C_{abcd} + 4 R^a_e C^{ebcd}C_{abcd} + \nabla_a J^a,
\end{align}
where $\nabla_a J^a$ is trivial anomaly. Following the same procedure, we get
\begin{align}
    a & = \frac13 L^5\pi^3k_{1,1},\\
    c_1 & = -\pi ^3 L^5 k_{1,1}+\frac{16}{3} \pi ^3 L^3 k_{2,3}+64 \pi ^3 L k_{3,7},\\
    c_2 & = -4 \pi ^3 L^5 k_{1,1}-\frac{64}{3} \pi ^3 L^3 k_{2,3}-64 \pi ^3 L k_{3,8},\\
    c_3 & = \frac{1}{3}L^5\pi^3 (k_{1,1}+ 16 \pi ^3 L^{-2} k_{2,3}) = \frac{\pi^6}{3024}\left.C_T\right|_{d=6}.
\end{align}
Then we have
\begin{equation}
    t_2  = \frac{15 \left(23 c_1-44 c_2+144 c_3\right)}{16 c_3},\qquad
    t_4  = -\frac{105 \left(c_1-2 c_2+6 c_3\right)}{2 c_3}\,,
\end{equation}
this means $(C_T, C_T t_2, C_T t_4)$ may be regarded as the arbitary even dimension generalization of the three $c$-charges. We may then recombine the Weyl invariants as follows
\begin{align}
    \tilde I^{(6)}_1 & = -\frac{16}{15}C^{abcd}C_{a}{}^{e}{}_{c}{}^{f}C_{bedf}+\frac{8}{15}C^{abcd}C_{ab}{}^{ef}C_{cdef},\nonumber\\
    \tilde I^{(6)}_2 & = -\frac{44}{105}C^{abcd}C_{a}{}^{e}{}_{c}{}^{f}C_{bedf}+\frac{23}{105}C^{abcd}C_{ab}{}^{ef}C_{cdef},\nonumber\\
    \tilde I^{(6)}_3 & = 96\nabla^{c}P^{ab}\nabla_{c}P_{ab}-96\nabla^{c}P^{ab}\nabla_{b}P_{ac}+3\nabla^{e}C^{abcd}\nabla_{e}C_{abcd}+96\nabla_{d}\nabla_{b}P_{ac}C^{abcd}\nonumber\\
    & \qquad +48P_{de}C^{abcd}C_{abc}{}^{e}-4C^{abcd}C_{a}{}^{e}{}_{c}{}^{f}C_{bedf}-7C^{abcd}C_{ab}{}^{ef}C_{cdef},
\end{align}
so that their $c$-charges becomes proportional to $C_T t_2$, $C_T t_4$, $C_T$ respectively
\begin{equation}
    \tilde c_1 = \tilde c_3t_2, \qquad \tilde c_2 = \tilde c_3 t_4, \qquad \tilde c_3 = c_3.
\end{equation}
It becomes clear that $\tilde I_1^{(6)}$ and $\tilde I^{(6)}_2$ are $W_{(3)}$ invariants, while $\tilde I^{(6)}_3$ is $W_{(2)}$ invariant. Thus the conjecture proposed in the main text holds for 6d holographic CFT from general higher curvature gravity.


\bibliography{refs}

\end{document}